\documentclass[aps, prb, reprint, amssymb, amsmath, superscriptaddress]{revtex4-1}

\usepackage{graphicx}
\usepackage{dcolumn}
\usepackage{bm}
\usepackage{amsmath}
\usepackage{amssymb}
\usepackage{latexsym}
\usepackage{epsfig}
\usepackage{amsbsy}
\usepackage{array}
\usepackage{amssymb}
\usepackage{setspace}
\usepackage{bm}

\def\sint{\ifmmode{- \!\!\!\!\!\! \int}
    \else{\hbox{$- \!\!\!\! \int \ $}}\fi}

\begin{document}


\title{Cooper Pairing in Insulating Valence Band in Fe-Based Superconductors}

\author{Lun-Hui Hu}
\affiliation{Department of Physics,  Zhejiang University, Hangzhou, China}
\affiliation{Collaborative Innovation Center of Advanced Microstructures, Nanjing, China}
\author{Wei-Qiang Chen}
\email{chenwq@sustc.edu.cn}
\affiliation{Department of Physics, South University of Sciences and Technology of China, Shenzhen, China}
\author{Fu-Chun Zhang}
\affiliation{Department of Physics,  Zhejiang University, Hangzhou, China}
\affiliation{Collaborative Innovation Center of Advanced Microstructures, Nanjing, China}

\date{\today}

\begin{abstract}
Conventional Cooper pairing arises from attractive interaction of electrons in the metallic bands. Recent experiment on Co-doped LiFeAs shows superconductivity in the insulating valence band, which is evolved from a metallic hole band upon doping.   Here we examine this phenomenon by studying  superconductivity in a three-orbital Hamiltonian relevant to the doped LiFeAs.  We show explicitly that Cooper pairing of the insulating hole band requires a
finite pairing interaction strength. For strong coupling, the superconductivity in the hole band is robust against the sink of the hole band below the Fermi level.  Our theory predicts a substantial upward shift of the chemical potential in the superconducting transition for Co-doped LiFeAs.
\end{abstract}



\maketitle

In conventional BCS theory, superconductivity arises as a Fermi surface instability, and
superconducting (SC) gap decreases exponentially with the density of states at the Fermi energy\cite{deGennes}.  So fully occupied or
empty bands are usually ignored in study of superconductivity.  It is interesting that the opening of a SC gap on a fully occupied band is observed in a recent ARPES experiment in a Fe-based superconductor \cite{Miao2014}.  Fe-based superconductors have generated great interest because they have highest transition temperature next to the cuprates at ambient pressure\cite{Kamihara2008, Chen2008-1,
  Chen2008-2, Wen2008, Chen2008, Ren2008-1, Ren2008-2, Ren2008-3}.  Typically, a Fe-based superconductor has a complex Fermi surface consisting of both electron pockets and hole pockets\cite{Lebegue2007, Singh2008, Xu2008, Cao2008, Zhang2009}.  In comparison with other
iron pnictides, LiFeAs  has a much shallower hole pocket around the center of the Brillouin
zone\cite{Borisenko2010, Eschrig2009, Lankau2010}.  Upon Co-doping, the entire hole band sinks below the Fermi energy, the hole pocket disappears and the hole band becomes insulating as revealed in the ARPES experiment.  Below the SC transition temperature $T_c$, the energy gap between the top of the hole band and the Fermi energy is found to be larger than the gap in the insulating state, which suggests Cooper pairing and superconductivity in the fully occupied hole band.\cite{Miao2014}

It is interesting to note that a similar phenomenon was discussed theoretically by Nozieres and Pistolesi\cite{Nozieres1999} over a decade ago. They considered superconductivity in a semiconductor with a small band gap, and found that the semiconductor may undergo a transition into SC state at low temperature if the
binding energy of the Cooper pair is larger than the energy cost to produce free charge carriers across the semiconducting gap.  LiFe$_{1-x}$Co$_x$ is a multi-band metal with electron band metallic and hole band fully occupied. The physics behind the superconductivity of the fully
occupied valence or hole band is similar to the superconductivity in a semiconductor.  In the multi-band Fe-based compound, electrons in the fully occupied hole band may be excited to the states  in metallic electron bands, which introduces an additional pairing channel.  This allows the realization of the band insulator to superconductor transition in multi-band LiFe$_{1-x}$Co$_x$.   In this paper, we consider a phenomenological mean field Hamiltonian based on a three-orbital model, relevant to LiFe$_{1-x}$Co$_x$, and study the Cooper pairing and superconductivity of the model.   We show explicitly that Cooper pairing in the insulating hole band requires a
finite pairing interaction strength. For strong pairing interaction, the superconductivity in the hole band is robust against the sink of the hole band below the Fermi level.  Our theory predicts a substantial upward shift of the chemical potential in the SC transition for Co-doped LiFeAs.

\begin{figure}[!htbp]
  \centering
  \begin{minipage}[b]{0.5\textwidth}
    \centering
    \includegraphics[width=3.2in]{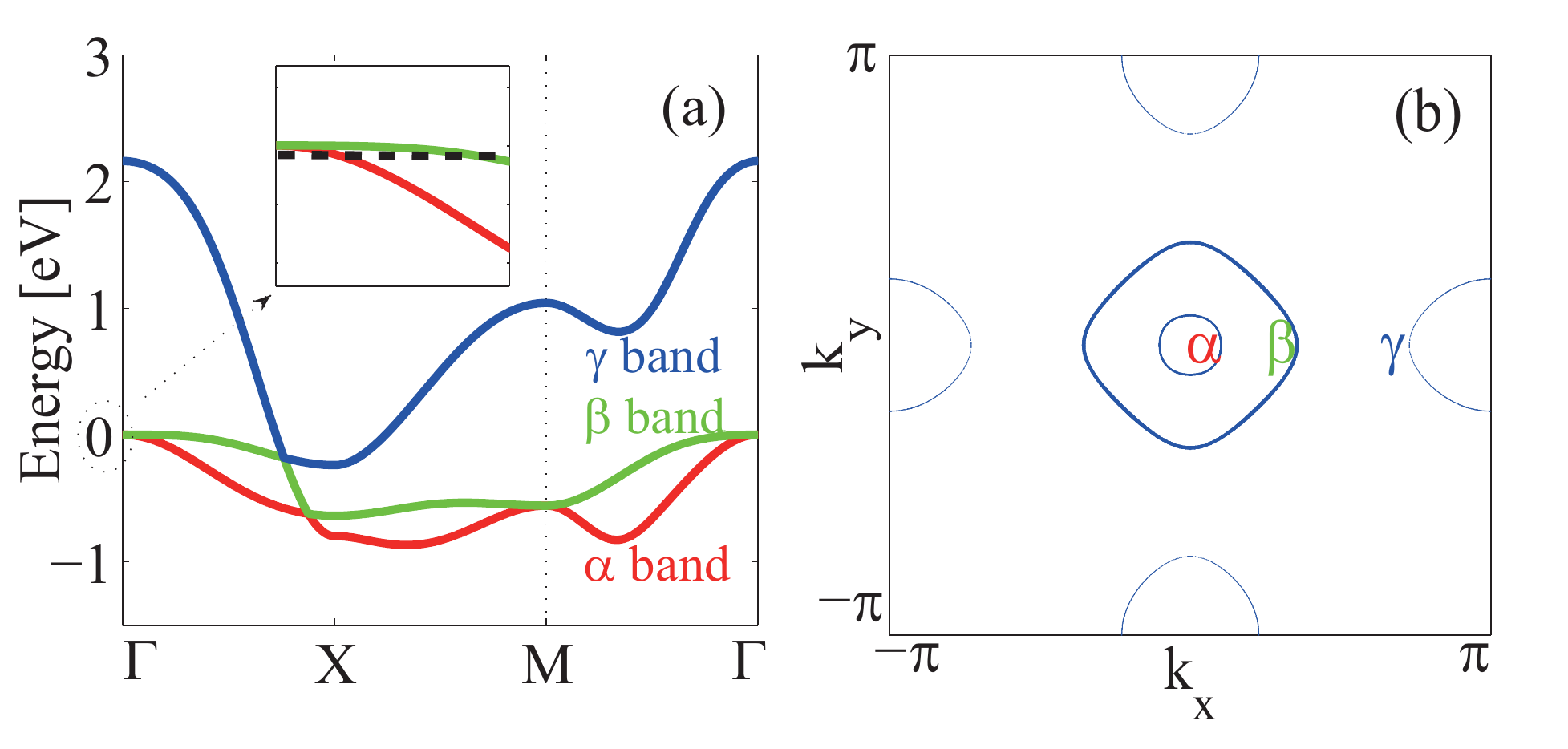}
  \end{minipage}
  \caption{(a): Band structure obtained from the tight binding model for LiFeAs. Red, green, and blue curves are for $\alpha$, $\beta$,
    and $\gamma$ band respectively. Black dashed line in the inset indicates the chemical potential in the absence of Co-doping.  (b): Fermi surfaces in the model with two hole pockets around \(\Gamma\)
    point and two electron pockets around \(X\) and \(Y\) point. In the calculations, the parameters in Eqn. (2) and (3) are chosen as, in units of eV,
    \(t_1=0.02, t_2=0.12, t_3=0.02, t_4=-0.046, t_5=0.2, t_6=0.3, t_7=-0.15, t_8=-t_7/2, t_9=-0.06, t_{10}=-0.03,
    t_{11}=0.014, \Delta_{xy}=1\),
    $\mu = 0.358$.
    \label{fig:band}}
\end{figure}

We consider the Hamiltonian
\begin{align}
\label{eq:1}
  \mathcal{H} & = \mathcal{H}_0 + \mathcal{H}_{pair},
\end{align}
where $H_0$ and $H_{pair}$ describes the kinetic part and the attractive interaction respectively.  In the LiFeAs
material, the states around the Fermi energy mainly consist of d$_{xz}$, d$_{yz}$, and d$_{xy}$ orbitals of Fe-3d
orbitals.  So we consider a three-orbital model introduced by Brydon et al.  \cite{Brydon2011}
\begin{align}
\label{eq:2}
  \mathcal{H}_0(\mathbf{k}) = \sum_{\mathbf{k}\sigma m n} T^{mn}(\mathbf{k}) c_{\mathbf{k}\sigma m}^{\dagger}
  c_{\mathbf{k}\sigma n} - \mu \sum_{\mathbf{k}\sigma m} c_{\mathbf{k}\sigma m}^{\dagger}
  c_{\mathbf{k}\sigma m}
\end{align}
where \(c_{\mathbf{k}\sigma m}^{\dagger}\)(\(c_{\mathbf{k}\sigma m}\))
creates (annihilates) an electron with orbital \(m\),
crystal wave-vector \(\mathbf{k}\)
and spin \(\sigma\),
$m, n = 1,2,3$ corresponds to d$_{xz}$, d$_{yz}$ and d$_{xy}$ orbital respectively.  The hopping integrals are given by
\begin{align}
\label{eq:3}
  T^{11} &=2t_1 \cos k_y + 2t_2\cos k_x + 4t_3\cos k_x\cos k_y \nonumber \\
  &\quad + 2t_{11}\left( \cos 2k_x-\cos 2k_y \right)  \nonumber \\
  T^{22} &=2t_1\cos k_x + 2t_2 \cos k_y + 4t_3\cos k_x\cos k_y \nonumber \\
  &\quad- 2t_{11}\left( \cos 2k_x-\cos 2k_y \right)  \nonumber \\
  T^{33} &=\Delta_{xy} + 2t_5 \left( \cos k_x + \cos k_y \right) + 4t_6 \cos k_x\cos k_y \nonumber \\
  &\quad + 2t_9\left( \cos 2k_x+\cos 2k_y \right) \nonumber \\
  &\quad + 4t_{10}\left( \cos 2k_x\cos k_y+\cos k_x\cos 2k_y \right) \nonumber \\
  T^{12} &=4t_4 \sin k_x\sin k_y   \nonumber \\
  T^{13} &=2it_7\sin k_x + 4it_8\sin k_x\cos k_y   \nonumber \\
  T^{23} &=2it_7\sin k_y + 4it_8\sin k_y\cos k_x
\end{align}
where $t_i$ are the hopping constants and chosen to be the same as in the paper by Brydon et al \cite{Brydon2011}.  The
band dispersions and the Fermi surfaces for LiFeAs are depicted in Fig.~\ref{fig:band}.  There are three bands, the
$\alpha$ band with smaller hole Fermi pocket, the $\beta$ band with larger hole Fermi pocket, and the $\gamma$ band with
electron Fermi pockets.  The chemical potential in the normal state is determined by the relative position of the top of
the $\alpha$ band.  We choose $\mu = 0.358 meV$ to model $LiFeAs$ in the absence of Co-doping. In this case, the chemical potential is $2 meV$ below the top of $\alpha$ band, consistent with the AREPS data.  The partial substitution of Co by Li introduces electrons into the FeAs plane, which
can be treated effectively as lifting the chemical potential \cite{Miao2014,Co1,Co2}.  In our calculations below, we choose $\mu= 0.364 meV$ and $\mu = 0.368 meV$ to model the Co-doped cases in experiments with the chemical potential $4 meV$  (1 per cent of Co-doping) and $8 meV$ (3 per cent of Co-doping) above the top of the $\alpha$ band, respectively.

 The band structure calculated in the tight binding model is in good agreement with the
 ARPES results.  However, we note that there is a degeneracy of the $\alpha$ and $\beta$ bands at the $\Gamma$ point in our model, which is absent in the ARPES data. Since the Co-doping mainly affects the $\alpha$ band but not much about the $\beta$ band as observed in the ARPES,  we shall drop out the $\beta$ band for simplicity in the calculations below to study the superconductivity of the insulating $\alpha$ band.  Namely, we will consider only the $\alpha$-hole band and the $\gamma$-
electron band.  The role of the $\beta$ band will be briefly discussed at the end of the paper and it is similar to the $\gamma$ band.

\begin{figure}[!htbp]
  \centering
  \begin{minipage}[b]{0.5\textwidth}
    \centering
    \includegraphics[width=3.2in]{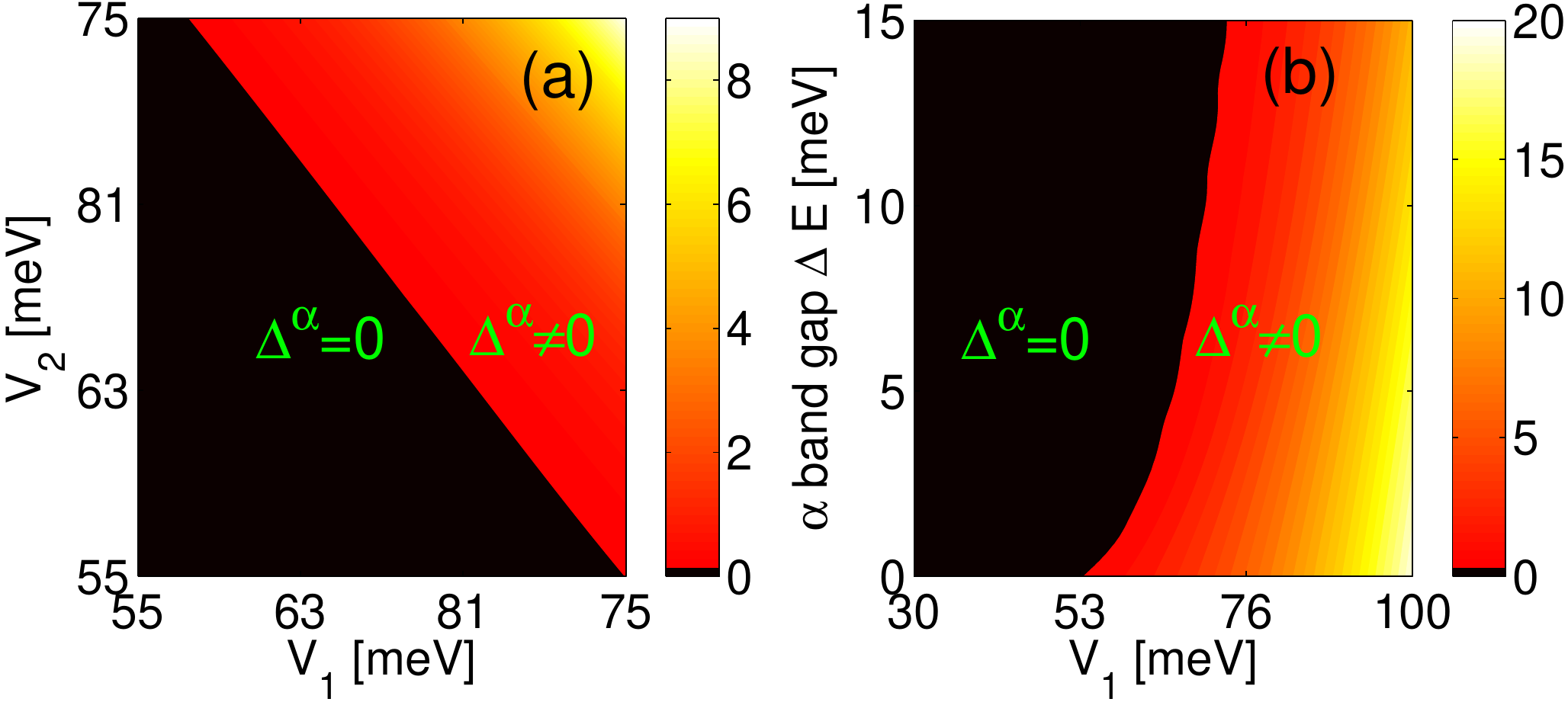}
  \end{minipage}
  \caption{\label{fig:phase_diagram} SC and insulating phase diagram of the valence $\alpha$ band at $T=0$, in parameter space $V_1$ and $V_2$ for a fixed $\Delta E = 4 meV$, (a); and in parameter space of $V_1$ and $\Delta E$ for a fixed $V_1/V_2 = 1.2$, (b).  $V_1$ and $V_2$ are intra-band pairing coupling and inter-band pair hopping, respectively, and $\Delta E$ is the $\alpha$-band gap, namely the energy difference between the top of the $\alpha$-band and the chemical potential $\mu$.  $\Delta E = 4 meV$ corresponds to Co-doping $x = 0.01$. In these calculations, $\mu$ is fixed. }
\end{figure}

We now consider superconductivity. For this purpose, we introduce a phenomenological pairing term $\mathcal{H}_{pair}$.  Because the $\alpha$ band and $\gamma$ band do not have any overlap near the Fermi energy,  we may safely neglect the inter-band pairings, and include the intra-band pairing coupling and the inter-band pair hopping process to study the superconductivity. We assume that the attractive interaction is from the next nearest neighbor intersite pairing \cite{JPHu}.  Then the gap function
in $\mathbf{k}$-space has the form  $\cos k_x \cos k_y$, and the pairing Hamiltonian reads
\begin{align}
\label{eq:4}
  \mathcal{H}_{\text{pair}} &= - \sum_{\mathbf{k},\tau=\alpha, \beta} \cos k_x \cos k_y \left( \Delta^{\tau} c_{ -\mathbf{k}
                              \downarrow}^{\tau \dag}c_{\mathbf{k}
                              \uparrow}^{\tau \dag}  + h.c. \right),
\end{align}
where $\Delta^{\tau}$ is the amplitude of the spin singlet SC gap in $\tau =\alpha, \gamma$ band,
 which are defined as
\begin{align}
  \label{eq:5}
  \Delta^{\alpha} &= \frac{4}{N}\sum_{\mathbf{k}}\cos k_x\cos k_y \left( V_1 \left\langle
                    c_{\mathbf{k} \uparrow}^{\alpha} c_{-\mathbf{k} \downarrow}^{\alpha} \right\rangle + V_2 \left\langle
                    c_{\mathbf{k} \uparrow}^{\gamma} c_{-\mathbf{k} \downarrow}^{\gamma} \right\rangle \right)
                    \nonumber \\
  \Delta^{\gamma} &= \frac{4}{N}\sum_{\mathbf{k}}\cos k_x\cos k_y \left( V_1 \left\langle
                    c_{\mathbf{k} \uparrow}^{\gamma} c_{-\mathbf{k} \downarrow}^{\gamma} \right\rangle + V_2 \left\langle
                    c_{\mathbf{k} \uparrow}^{\alpha} c_{-\mathbf{k} \downarrow}^{\alpha} \right\rangle \right).
\end{align}
The self-consistent equations then read
\begin{align}
\label{eq:8}
  \Delta^{\alpha(\gamma)} &= \frac{4}{N} \sum_{\mathbf{k}}[V_1\Delta^{\alpha(\gamma)} + V_2 \Delta^{\gamma(\alpha)}] \frac{\cos^2 k_x\cos^2
                            k_y}{E_{\mathbf{k}}^{\alpha(\gamma)}}
                            \nonumber \\
                          & \phantom{\frac{4}{N} \sum_{\mathbf{k}}}\times \tanh \left[ E_{\mathbf{k}}^{\alpha(\gamma)}/2 k_B T \right]
\end{align}
where
$E_{\mathbf{k}}^{ \alpha(\gamma)}=\sqrt{\left[\epsilon_{\mathbf{k}}^{ \alpha(\gamma)}-\mu\right]^2+ \left[\Delta^{\alpha
      (\gamma)}\cos k_x\cos k_y\right]^2}$
and $\epsilon_{\mathbf{k}}^{\alpha(\gamma)}$ are the quasiparticle energy and single particle energy of
$\alpha$($\gamma$) band, respectively.

\begin{figure}[!htbp]
  \centering
  \begin{minipage}[b]{0.5\textwidth}
    \includegraphics[width=3.2in]{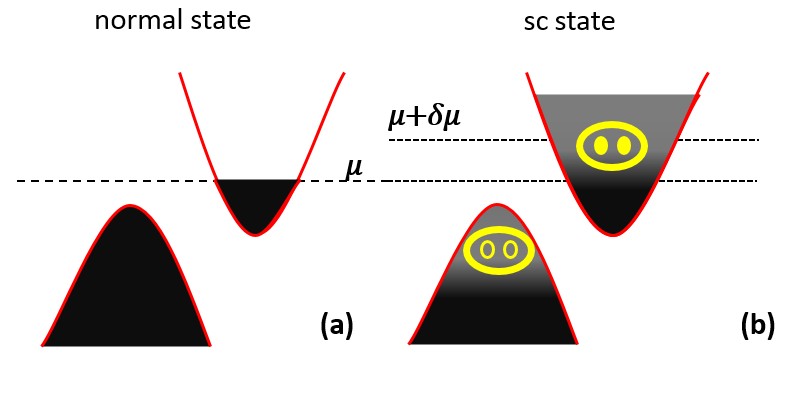}\\
    \includegraphics[width=3.2in]{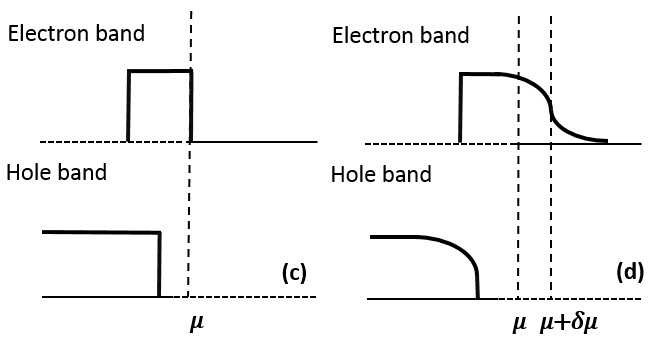}
  \end{minipage}
  \caption{Schematic illustration of the electronic bands, panels (a) and (b),  and the mean occupation number of the
    electrons of a state, panel (c) and (d), in the normal and SC states. In the SC transition, some of electron pairs in the fully occupied valence band transfer to the conduction band, which leads to Cooper pairing in the insulating hole band.  Chemical potential $\mu$ in the normal state is shifted upward to $\mu+ \delta \mu$ in the SC state due to the transferred electrons into the conduction band.   \label{fig:sketch}}
\end{figure}

 We first discuss the zero temperature ($T=0$) case.  The superconductivity in the $\alpha$ band is controlled by $V_1$, $V_2$, and the $\alpha$ band gap $\Delta E$.  For simplicity, we choose a gauge where $\Delta^{\alpha}$ is
real and non-negative.  By solving the self-consistent equation for various $V_1$ and $V_2$ at a fixed $\Delta E=4 meV$, we find a phase diagram plotted in Fig. \ref{fig:phase_diagram}(a).  The SC gap in the $\alpha$ band, $\Delta^{\alpha}$ increases with the increase of the coupling constant, and
$V_1$ and $V_2$ play a similar role in enhancing the superconductivity.  The dependence of the $\Delta^{\alpha}$ on $\Delta E$ and $V_1$ is plotted in Fig. \ref{fig:phase_diagram}(b) for a fixed value of $V_1/V_2=1.2$ .  From Fig. 3, it is clear that the insulating $\alpha$-band becomes SC only at $V_1> V_c$ with $V_c$ the critical pairing coupling. This conclusion is similar to the study on semiconductor pairings \cite{Nozieres1999}.  $\Delta^\alpha$ reduces as  $V_1$ reduces to $V_{c}$, and vanishes
when $V_1 < V_{c}$.  In conventional BCS theory, the superconductivity instability is induced by infinitesimal coupling constant.  For an insulating valence band,
one must remove some electrons from the fully occupied valence band to the conduction band, which will cost an energy at least equal to the insulating band gap $\Delta E$.  Therefore, the $\alpha$ band may become SC only when the energy gain by forming a
Cooper pair is larger than the energy cost, which leads to a finite critical $V_{c}$.

\begin{figure}[htbp]
  \centering
  \begin{minipage}[b]{0.5\textwidth}
    \centering
    \includegraphics[width=3.2in]{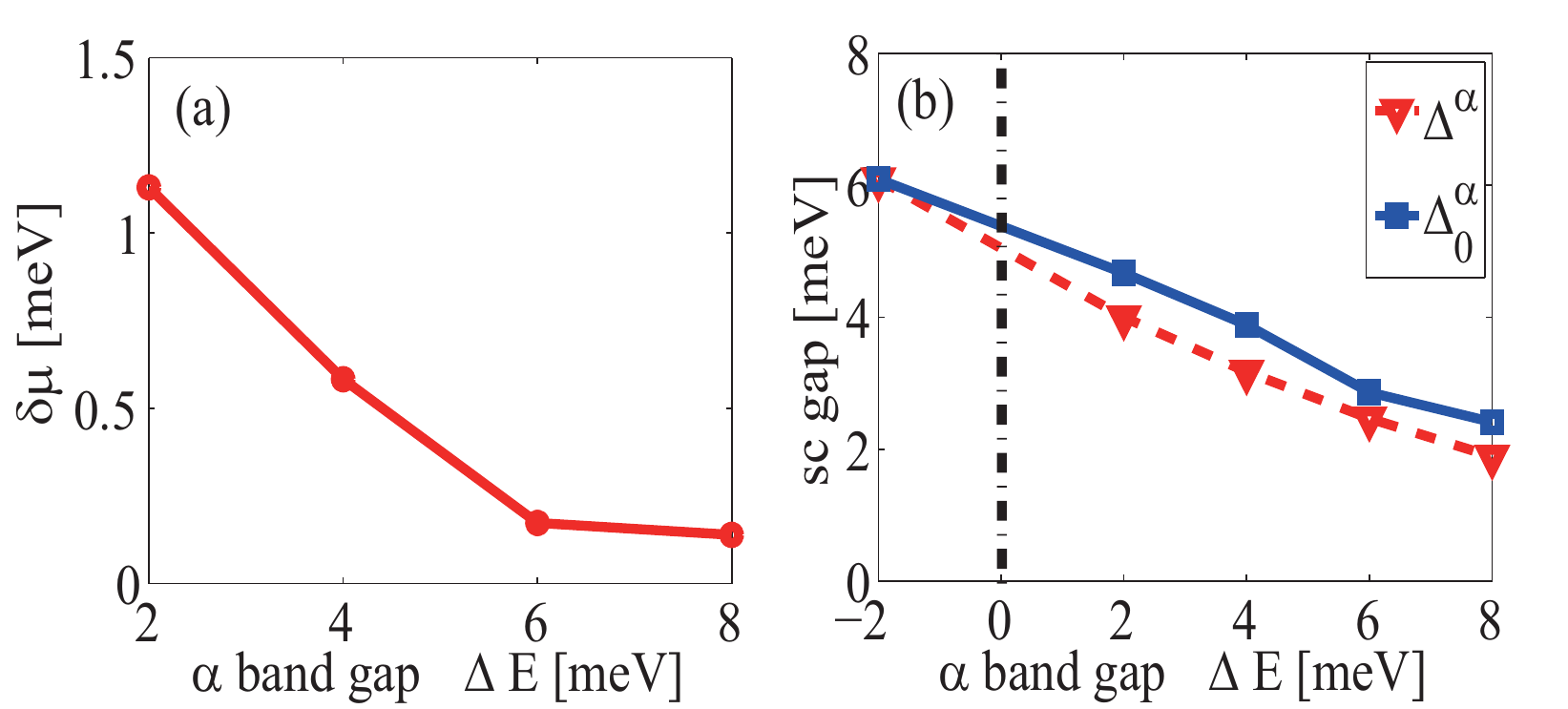}
  \end{minipage}
  \caption[]{\label{fig:mu_shift} Chemical potential shift $\delta \mu$ in the SC state as function of the insulating $\alpha$ band gap $\Delta E$ (panel a); and the SC gaps on the $\alpha$ band with $\delta \mu$ included (red dashed curve) and neglected (blue solid curve), and the SC gap on $\gamma$-band (green dashed line), (panel b). The parameters are $V_1 = 76 meV$ and $V_1 / V_2 = 1.2$}
\end{figure}

In the above calculations, the chemical potential $\mu$ is fixed.  In the conventional BCS theory for metallic normal
state, the chemical potential shift is tiny if we fix the total electron density.  The situation is different in the
LiFe$_{1-x}$Co$_x$As case.  As we analyzed above, some electrons must be removed from hole band to induce
superconductivity in the $\alpha$ band.  If we consider the total electron density is fixed as in the usual case, the
only place those electron can go is the $\gamma$ band.  That increases the number of electrons in the electron band and
leads to an upward shift of the chemical potential as shown in fig.~\ref{fig:sketch}.  The chemical potential shift may
be calculated through the equation for number of electrons per site
\begin{align}
\label{eq:9}
  n &=\frac{1}{N} \sum_{\mathbf{k}}\left\lbrack \left( 1-\frac{\epsilon_{\mathbf{k}}^{\alpha}-\mu}{E_{\mathbf{k}}^{\alpha}}
      \right) + \left(1-\frac{\epsilon_{\mathbf{k}}^{\gamma}-\mu}{E_{\mathbf{k}}^{\gamma}}\right)  \right\rbrack.
\end{align}
We have solved eqn.~\eqref{eq:8} and eqn.~\eqref{eq:9} for the couplings $V_1 = 76 meV$ and
$V_1 / V_2 = 1.2$ at various $\alpha$ band gap $\Delta E$.  The qualitative behavior on the superconductivity is similar to that we discussed at the fixed chemical potential, and the
SC gap reduces with increasing $\alpha$ band gap as shown in fig.~\ref{fig:mu_shift}(b).  For the present model, as
$\Delta E$ increases from 4 meV to 8 meV,
$\Delta^{\alpha}$ reduces from 6 meV to 1.8 meV.  The calculated chemical potential shift $\delta \mu$ is depicted
in fig.~\ref{fig:mu_shift}(a), which is rather large
especially in the case with larger $\Delta^{\alpha}$.  For example, $\delta \mu \sim 1.1 meV$ with $\Delta E = 2 meV$,
while the SC gap on $\alpha$ band is just around $3.9 meV$.  Therefore, the chemical potential shift can not be
simply neglected in this case.  Similar with the SC gap, the chemical potential shift also reduces with
increasing the $\alpha$ band gap $\Delta E$.  The reason is that the total number of electrons excited from $\alpha$
band to $\gamma$ band reduces with suppression of the SC gap, which leads to the reduction of $\delta \mu$.

The chemical potential shift $\delta \mu$ may be important for the estimation of the SC gap from the experimental data.  If one neglects $\delta \mu$, the SC gap on the $\alpha$ band can be calculated from the gap at $\Gamma$
point in the normal state ($\Delta E$) and gap in the SC state ($\Delta_{\text{after sc}}$),
\begin{align}
\label{eq:10}
  \Delta^{\alpha}_0 = \sqrt{\Delta_{\text{after sc}}^2-\Delta E^2}
\end{align}
Including $\delta \mu$ , the SC gap is given by
\begin{align}
\label{eq:11}
  \Delta^{\alpha} = \sqrt{\Delta_{\text{after sc}}^2-(\Delta E + \delta \mu)^2}
\end{align}
From the above analyses, the SC gap estimated without including the chemical potential shift is larger than the correct one.  In fig.~\ref{fig:mu_shift}(b), we depict the value of
$\Delta_0^{\alpha}$ (blue solid line) and the correct SC gap $\Delta^\alpha$ (red dashed line) for comparison.

\begin{figure}[htbp]
  \centering
  \begin{minipage}[b]{0.5\textwidth}
    \centering
    \includegraphics[width=1.6in]{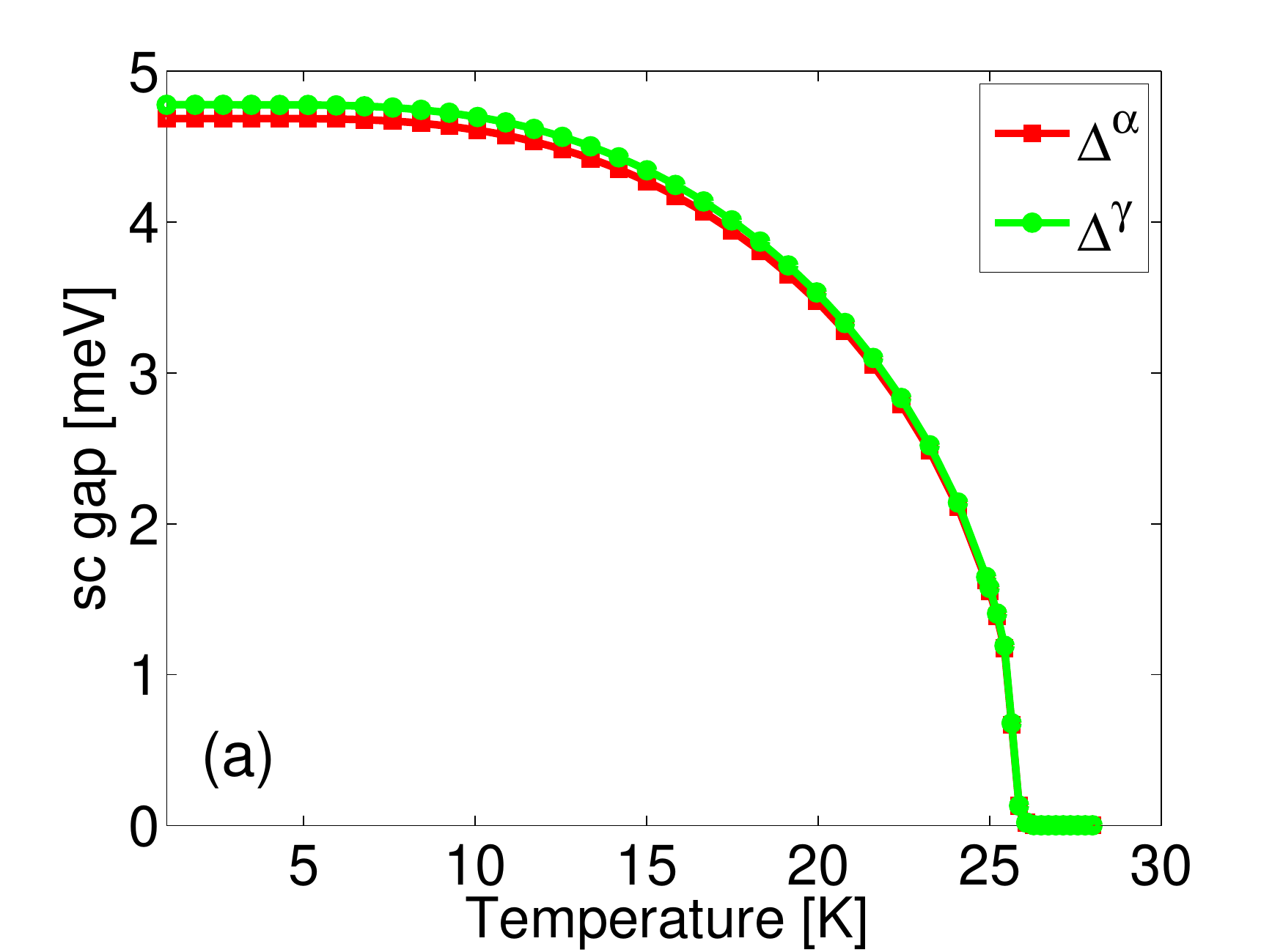}
    \includegraphics[width=1.6in]{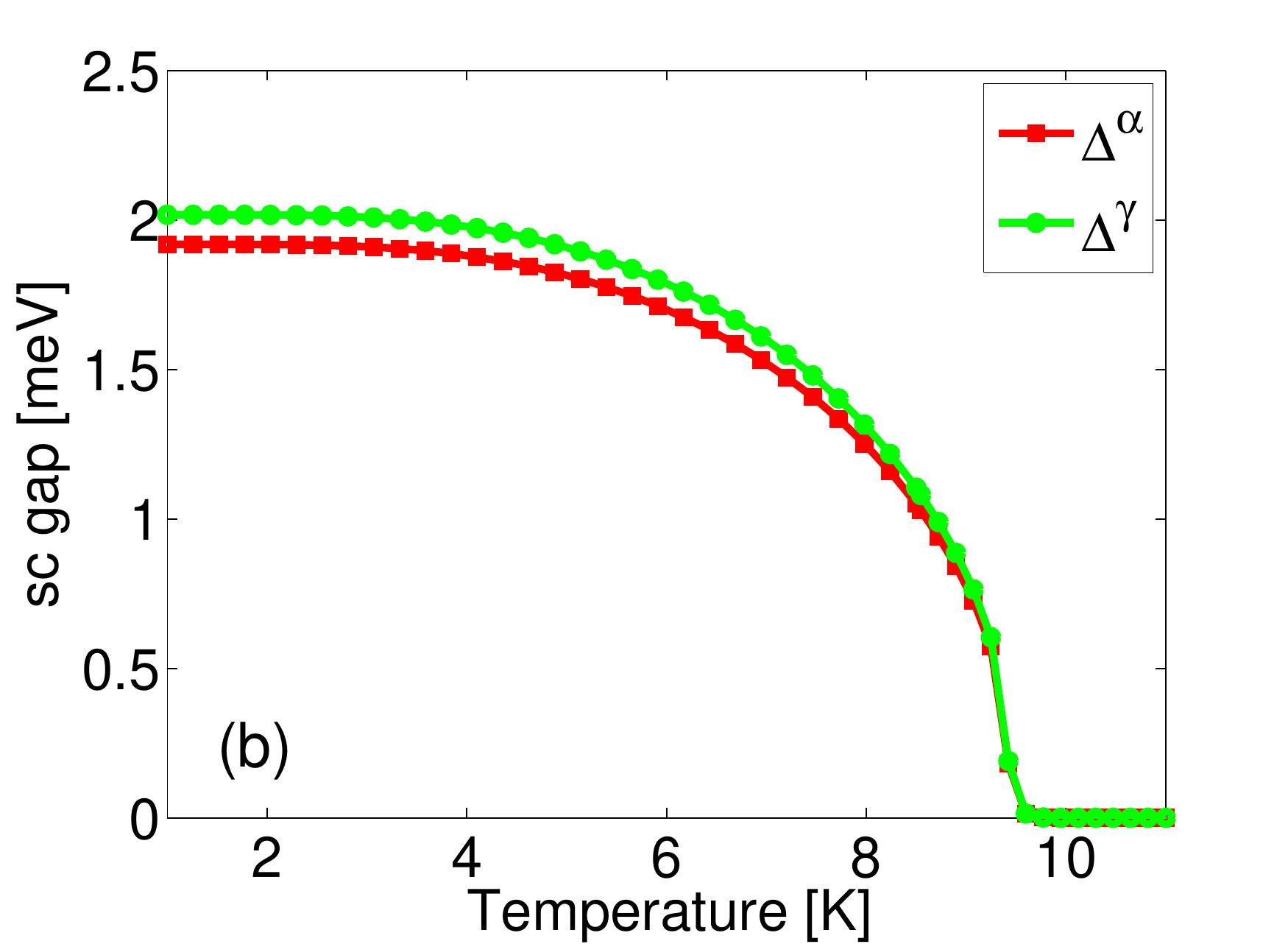}
    \includegraphics[width=1.6in]{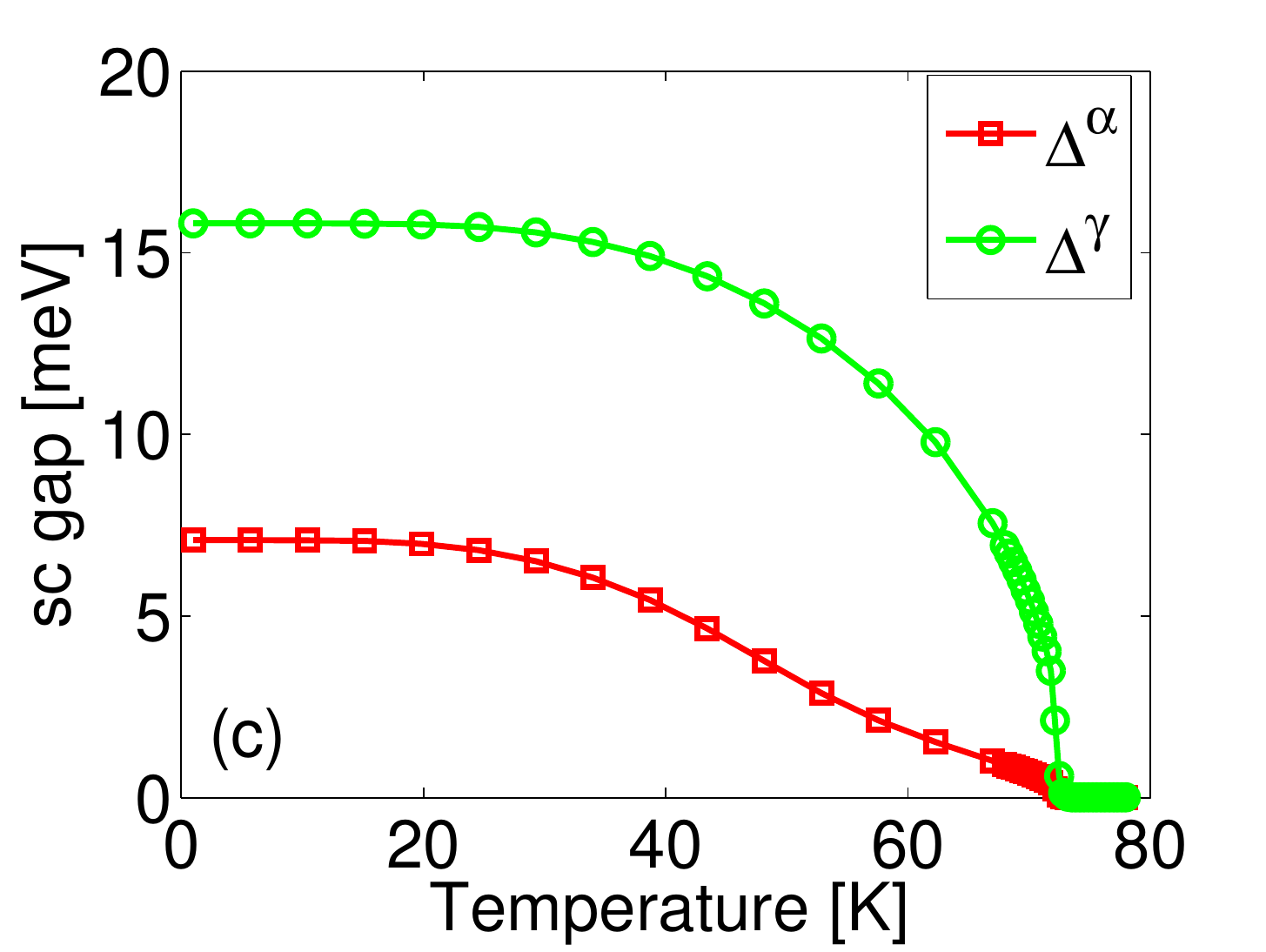}
    \includegraphics[width=1.6in]{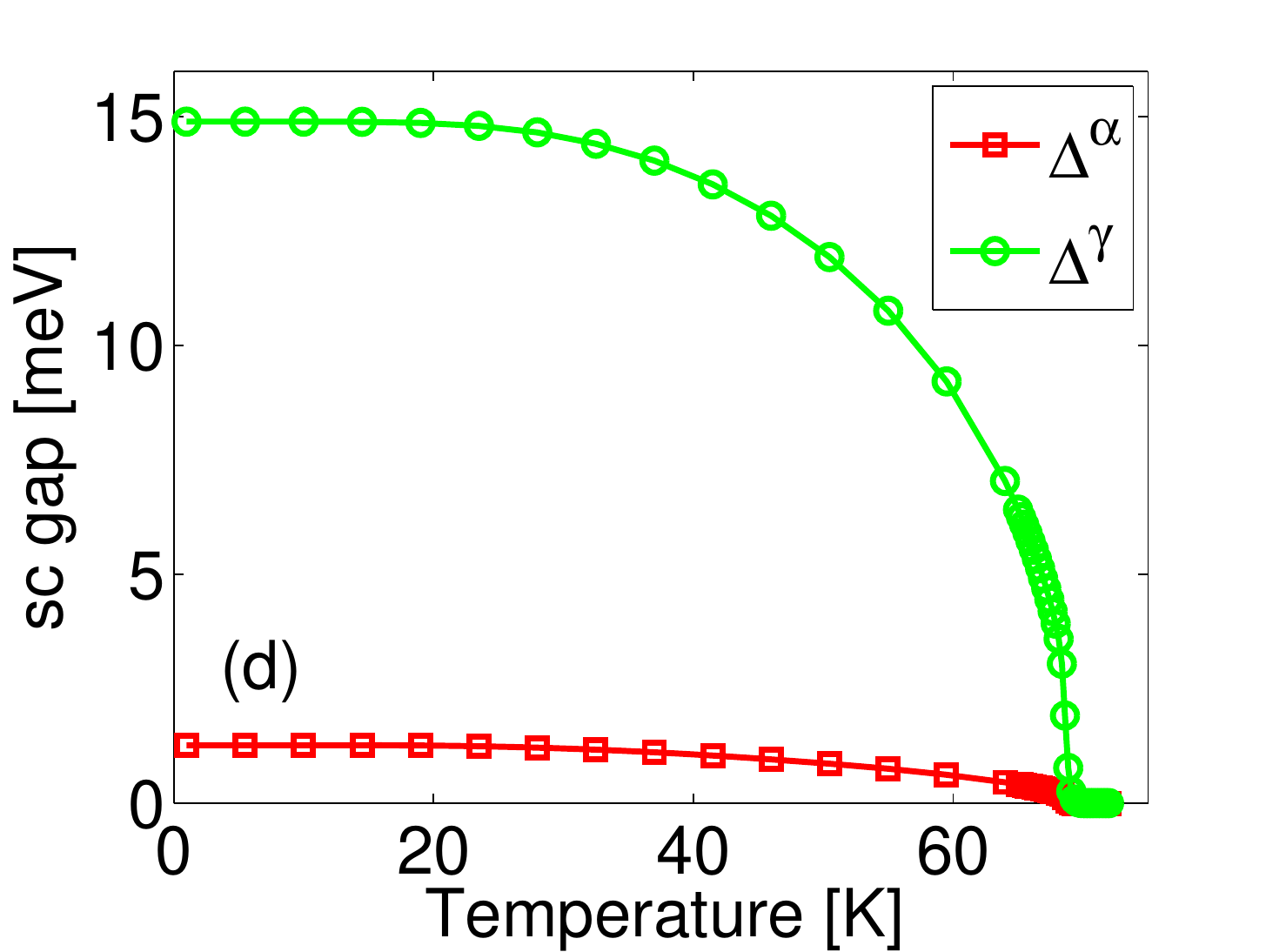}
  \end{minipage}
  \caption[]{\label{fig:finite_T} Temperature dependence of SC gap on $\alpha$ and $\gamma$ band from model calculations.  In upper panel, $V_1 = 76 meV$ , $V_1 / V_2 = 1.2$.  (a): $\Delta E = 2 meV$; and  (b): $\Delta E = 8 meV$. In lower panel,  $ V_1 = 160 meV$,  $V_2 = 1 meV$. (c): $\Delta E = 2
    meV$ ;  (d):  $\Delta E = 8 meV$.}
\end{figure}

We now study superconductivity at finite temperatures.  The SC gap is reduced with the increase of temperature as shown in fig.~\ref{fig:finite_T},
where the results for various $V_1$, $V_2$, and $\Delta E$ are depicted.  As shown in the figure,
because of the presence of interband pair hopping, the SC gap of $\alpha$ band and $\gamma$ band vanishes
at the same temperature Tc though they are different at zero temperature.  If one sets $V_2 = 0$, the $\alpha$ band and $\gamma$ band are decoupled in the SC transition, and they may have different $T_{c}$.

In the above calculations, we have neglected the $\beta$ hole band.  The $\beta$ band will play a similar role as the electron
band to the $\alpha$ band superconductivity.  The $\beta$ band will enhance the superconductivity on the $\alpha$ band through the pair hopping between the two bands.  The effect to the chemical potential shift is mixed.  The enhancement of superconductivity will increase the number of electrons in the $\alpha$ band to be excited, hence to increase the chemical potential shift upward.  On the other hand, the $\beta$ band provides additional states for
the $\alpha$ electrons to transfer to, which will increase the density of states at Fermi energy and reduce the chemical potential shift.  So the quantitative result depends on the parameters.  But the qualitative result will not
change.

In summary, we have presented a theory to explain the observed superconductivity in the $\alpha$-hole band which is completely below the Fermi energy upon Co-doping in the LiFe$_{1-x}$Co$_x$As.  We consider a three-orbital
Hamiltonian with an intra-band spin-singlet pairing and an inter-band pair hopping terms.  The $\alpha$ band is shown to become superconducting if and only if the pairing strength exceeds a critical value. Therefore the observation of superconductivity in an insulating valence band is an experimental proof of the strong coupling in Fe-based superconductors.
For strong pairing interaction, the superconductivity in the hole band is robust against the band gap of the hole band.  We show that the chemical potential shift accompanied with the
superconducting transition is rather large and should be taken into account in extracting the superconducting gap from the ARPES experiment.

This work is partially supported by NSFC projects 11204186 and 11374135 (WQC) and 11274269 (FCZ).


\begin{thebibliography}{35}
\expandafter\ifx\csname
natexlab\endcsname\relax\def\natexlab#1{#1}\fi
\expandafter\ifx\csname bibnamefont\endcsname\relax
  \def\bibnamefont#1{#1}\fi
\expandafter\ifx\csname bibfnamefont\endcsname\relax
  \def\bibfnamefont#1{#1}\fi
\expandafter\ifx\csname citenamefont\endcsname\relax
  \def\citenamefont#1{#1}\fi
\expandafter\ifx\csname url\endcsname\relax
  \def\url#1{\texttt{#1}}\fi
\expandafter\ifx\csname
urlprefix\endcsname\relax\def\urlprefix{URL }\fi
\providecommand{\bibinfo}[2]{#2}
\providecommand{\eprint}[2][]{\url{#2}}

\bibitem{deGennes} P.G. de Gennes, Superconductivity of Metals and Alloys (Addison-Wesley, New York, 1989).

\bibitem[{\citenamefont{Miao et~al.}(2014)\citenamefont{Miao, Qian, Shi, Richard, Kim, Hoesch, Xing, Wang, Jin, Hu, Ding}}]{Miao2014}
\bibinfo{author}{\bibfnamefont{H.}~\bibnamefont{Miao}},
\bibinfo{author}{\bibfnamefont{T.}~\bibnamefont{Qian}},
\bibinfo{author}{\bibfnamefont{X.}~\bibnamefont{Shi}},
\bibinfo{author}{\bibfnamefont{P.}~\bibnamefont{Richard}},
\bibinfo{author}{\bibfnamefont{T.~K.}~\bibnamefont{Kim}},
\bibinfo{author}{\bibfnamefont{M.}~\bibnamefont{Hoesch}},
\bibinfo{author}{\bibfnamefont{L.~Y.}~\bibnamefont{Xing}},
\bibinfo{author}{\bibfnamefont{X.~C.}~\bibnamefont{Wang}},
\bibinfo{author}{\bibfnamefont{C.~Q.}~\bibnamefont{Jin}},
\bibinfo{author}{\bibfnamefont{J.~P.}~\bibnamefont{Hu}}
\bibnamefont{and} \bibinfo{author}{\bibfnamefont{H.}~\bibnamefont{Ding}},
\bibinfo{journal}{Nat. Commun}
\textbf{\bibinfo{volume}{6}}, \bibinfo{pages}{6056}
(\bibinfo{year}{2015}).


\bibitem[{\citenamefont{Kamihara et~al.}(2008)\citenamefont{Kamihara, Watanabe, Hirano, Hosono}}]{Kamihara2008}
\bibinfo{author}{\bibfnamefont{Y.}~\bibnamefont{Kamihara}},
\bibinfo{author}{\bibfnamefont{T.}~\bibnamefont{Watanabe}},
\bibinfo{author}{\bibfnamefont{M.}~\bibnamefont{Hirano}}
\bibnamefont{and} \bibinfo{author}{\bibfnamefont{H.}~\bibnamefont{Hosono}},
\bibinfo{journal}{Am. Chem. Soc.}
\textbf{\bibinfo{volume}{130}}, \bibinfo{pages}{3296}
(\bibinfo{year}{2008}).

\bibitem[{\citenamefont{Chen et~al.}(2008)\citenamefont{Chen, Li, Li, Zhou, Wu, Dong, Hu, Zheng, Chen, Yuan, Singleton, Luo, Wang}}]{Chen2008-1}
\bibinfo{author}{\bibfnamefont{G.~F.}~\bibnamefont{Chen}},
\bibinfo{author}{\bibfnamefont{Z.}~\bibnamefont{Li}},
\bibinfo{author}{\bibfnamefont{G.}~\bibnamefont{Li}},
\bibinfo{author}{\bibfnamefont{J.}~\bibnamefont{Zhou}},
\bibinfo{author}{\bibfnamefont{D.}~\bibnamefont{Wu}},
\bibinfo{author}{\bibfnamefont{J.}~\bibnamefont{Dong}},
\bibinfo{author}{\bibfnamefont{W. Z.}~\bibnamefont{Hu}},
\bibinfo{author}{\bibfnamefont{P.}~\bibnamefont{Zheng}},
\bibinfo{author}{\bibfnamefont{Z.~J.}~\bibnamefont{Chen}},
\bibinfo{author}{\bibfnamefont{H.~Q.}~\bibnamefont{Yuan}},
\bibinfo{author}{\bibfnamefont{J.}~\bibnamefont{Singleton}},
\bibinfo{author}{\bibfnamefont{J.~L.}~\bibnamefont{Luo}}
\bibnamefont{and} \bibinfo{author}{\bibfnamefont{N.~L.}~\bibnamefont{Wang}},
\bibinfo{journal}{Phys. Rev. Lett.}
\textbf{\bibinfo{volume}{101}}, \bibinfo{pages}{057007}
(\bibinfo{year}{2008}).

\bibitem[{\citenamefont{Chen et~al.}(2008)\citenamefont{Chen, Li, Wu, Li, Hu, Dong, Zheng, Luo, Wang}}]{Chen2008-2}
\bibinfo{author}{\bibfnamefont{G.~F.}~\bibnamefont{Chen}},
\bibinfo{author}{\bibfnamefont{Z.}~\bibnamefont{Li}},
\bibinfo{author}{\bibfnamefont{D.}~\bibnamefont{Wu}},
\bibinfo{author}{\bibfnamefont{G.}~\bibnamefont{Li}},
\bibinfo{author}{\bibfnamefont{W. Z.}~\bibnamefont{Hu}},
\bibinfo{author}{\bibfnamefont{J.}~\bibnamefont{Dong}},
\bibinfo{author}{\bibfnamefont{P.}~\bibnamefont{Zheng}},
\bibinfo{author}{\bibfnamefont{J.~L.}~\bibnamefont{Luo}}
\bibnamefont{and} \bibinfo{author}{\bibfnamefont{N.~L.}~\bibnamefont{Wang}},
\bibinfo{journal}{Phys. Rev. Lett.}
\textbf{\bibinfo{volume}{100}}, \bibinfo{pages}{247002}
(\bibinfo{year}{2008}).


\bibitem[{\citenamefont{Wen et~al.}(2008)\citenamefont{Wen, Mu, Fang, Yang, Zhu}}]{Wen2008}
\bibinfo{author}{\bibfnamefont{H.-H.}~\bibnamefont{Wen}},
\bibinfo{author}{\bibfnamefont{G.}~\bibnamefont{Mu}},
\bibinfo{author}{\bibfnamefont{L.}~\bibnamefont{Fang}},
\bibinfo{author}{\bibfnamefont{H.}~\bibnamefont{Yang}}
\bibnamefont{and} \bibinfo{author}{\bibfnamefont{X.}~\bibnamefont{Zhu}},
\bibinfo{journal}{Europhys. Lett.}
\textbf{\bibinfo{volumne}{82}}, \bibinfo{pages}{17009}
(\bibinfo{year}{2008}).


\bibitem[{\citenamefont{Chen et~al.}(2008)\citenamefont{Chen, Wu, Wu, Liu, Chen, Fang}}]{Chen2008}
\bibinfo{author}{\bibfnamefont{X.~H.}~\bibnamefont{Chen}},
\bibinfo{author}{\bibfnamefont{T.}~\bibnamefont{Wu}},
\bibinfo{author}{\bibfnamefont{G.}~\bibnamefont{Wu}},
\bibinfo{author}{\bibfnamefont{R.~H.}~\bibnamefont{Liu}},
\bibinfo{author}{\bibfnamefont{H.}~\bibnamefont{Chen}}
\bibnamefont{and} \bibinfo{author}{\bibfnamefont{D.~F.}~\bibnamefont{Fang}},
\bibinfo{journal}{Nature}
\textbf{\bibinfo{volume}{453}}, \bibinfo{pages}{761}
(\bibinfo{year}{2008}).


\bibitem[{\citenamefont{Ren et~al.}(2008)\citenamefont{Ren, Yang, Lu, Yi, Che, Dong, Sun, Zhao}}]{Ren2008-1}
\bibinfo{author}{\bibfnamefont{Z.~A.}~\bibnamefont{Ren}},
\bibinfo{author}{\bibfnamefont{J.}~\bibnamefont{Yang}},
\bibinfo{author}{\bibfnamefont{W.}~\bibnamefont{Lu}},
\bibinfo{author}{\bibfnamefont{W.}~\bibnamefont{Yi}},
\bibinfo{author}{\bibfnamefont{G.}~\bibnamefont{Che}},
\bibinfo{author}{\bibfnamefont{X.}~\bibnamefont{Dong}},
\bibinfo{author}{\bibfnamefont{L.}~\bibnamefont{Sun}}
\bibnamefont{and} \bibinfo{author}{\bibfnamefont{Z.}~\bibnamefont{Zhao}}
\bibinfo{journal}{Mater. Res. Innovat.}
\textbf{\bibinfo{volume}{12}}, \bibinfo{pages}{105}
(\bibinfo{year}{2008}).


\bibitem[{\citenamefont{Ren et~al.}(2008)\citenamefont{Ren, Lu, Yang, Yi, Shen, Li, Che, Dong, Sun, Zhou, Zhao}}]{Ren2008-2}
\bibinfo{author}{\bibfnamefont{Z.~A.}~\bibnamefont{Ren}},
\bibinfo{author}{\bibfnamefont{W.}~\bibnamefont{Lu}},
\bibinfo{author}{\bibfnamefont{J.}~\bibnamefont{Yang}},
\bibinfo{author}{\bibfnamefont{W.}~\bibnamefont{Yi}},
\bibinfo{author}{\bibfnamefont{X.~L.}~\bibnamefont{Shen}},
\bibinfo{author}{\bibfnamefont{Z.~C.}~\bibnamefont{Li}},
\bibinfo{author}{\bibfnamefont{G.~C.}~\bibnamefont{Che}},
\bibinfo{author}{\bibfnamefont{X.~L.}~\bibnamefont{Dong}},
\bibinfo{author}{\bibfnamefont{L.~L.}~\bibnamefont{Sun}}
\bibinfo{author}{\bibfnamefont{F.}~\bibnamefont{Zhou}}
\bibnamefont{and} \bibinfo{author}{\bibfnamefont{Z.}~\bibnamefont{Zhao}}
\bibinfo{journal}{Chin. Phys. Lett.}
\textbf{\bibinfo{volume}{25}}, \bibinfo{pages}{2215}
(\bibinfo{year}{2008}).


\bibitem[{\citenamefont{Ren et~al.}(2008)\citenamefont{Ren, Che, Dong, Yang, Lu, Yi, Shen, Li, Sun, Zhou, Zhao}}]{Ren2008-3}
\bibinfo{author}{\bibfnamefont{Z.~A.}~\bibnamefont{Ren}},
\bibinfo{author}{\bibfnamefont{G.~C.}~\bibnamefont{Che}},
\bibinfo{author}{\bibfnamefont{X.~L.}~\bibnamefont{Dong}},
\bibinfo{author}{\bibfnamefont{J.}~\bibnamefont{Yang}},
\bibinfo{author}{\bibfnamefont{W.}~\bibnamefont{Lu}},
\bibinfo{author}{\bibfnamefont{W.}~\bibnamefont{Yi}},
\bibinfo{author}{\bibfnamefont{X.~L.}~\bibnamefont{Shen}},
\bibinfo{author}{\bibfnamefont{Z.~C.}~\bibnamefont{Li}},
\bibinfo{author}{\bibfnamefont{L.~L.}~\bibnamefont{Sun}},
\bibinfo{author}{\bibfnamefont{F.}~\bibnamefont{Zhou}}
\bibnamefont{and} \bibinfo{author}{\bibfnamefont{Z.}~\bibnamefont{Zhao}}
\bibinfo{journal}{Europhys. Lett.}
\textbf{\bibinfo{volume}{83}}, \bibinfo{pages}{17002}
(\bibinfo{year}{2008}).

\bibitem[{\citenamefont{Lebegue.}(2007)\citenamefont{Lebegue}}]{Lebegue2007}
\bibinfo{author}{\bibfnamefont{S.}~\bibnamefont{Lebegue}},
\bibinfo{journal}{Phys. Rev. B}
\textbf{\bibinfo{volume}{75}}, \bibinfo{pages}{035110}
(\bibinfo{year}{2007}).


\bibitem[{\citenamefont{Singh et~al.}(2008)\citenamefont{Singh, Du}}]{Singh2008}
\bibinfo{author}{\bibfnamefont{D.~J.}~\bibnamefont{Singh}}
\bibnamefont{and} \bibinfo{author}{\bibfnamefont{M.~H.}~\bibnamefont{Du}}
\bibinfo{journal}{Phys. Rev. Lett.}
\textbf{\bibinfo{volume}{100}}, \bibinfo{pages}{237003}
(\bibinfo{year}{2008}).


\bibitem[{\citenamefont{Xu et~al.}(2008)\citenamefont{Xu, Ming, Yao, Dai, Zhang, Fang}}]{Xu2008}
\bibinfo{author}{\bibfnamefont{G.}~\bibnamefont{Xu}},
\bibinfo{author}{\bibfnamefont{W.}~\bibnamefont{Ming}},
\bibinfo{author}{\bibfnamefont{Y.}~\bibnamefont{Yao}},
\bibinfo{author}{\bibfnamefont{X.}~\bibnamefont{Dai}},
\bibinfo{author}{\bibfnamefont{S.~C.}~\bibnamefont{Zhang}}
\bibnamefont{and} \bibinfo{author}{\bibfnamefont{Z.}~\bibnamefont{Fang}}
\bibinfo{journal}{Europhys. Lett.}
\textbf{\bibinfo{volume}{82}}, \bibinfo{pages}{67002}
(\bibinfo{year}{2008}).


\bibitem[{\citenamefont{Cao et~al.}(2008)\citenamefont{Cao, Hirschfeld, Cheng}}]{Cao2008}
\bibinfo{author}{\bibfnamefont{C.}~\bibnamefont{Cao}},
\bibinfo{author}{\bibfnamefont{P.~J.}~\bibnamefont{Hirschfeld}}
\bibnamefont{and} \bibinfo{author}{\bibfnamefont{H.~P.}~\bibnamefont{cheng}}
\bibinfo{journal}{Phys. Rev. B}
\textbf{\bibinfo{volume}{77}}, \bibinfo{pages}{220506}
(\bibinfo{year}{2008}).

\bibitem[{\citenamefont{Zhang et~al.}(2009)\citenamefont{Zhang, Xu, Dai, Fang}}]{Zhang2009}
\bibinfo{author}{\bibfnamefont{H.~J.}~\bibnamefont{Zhang}},
\bibinfo{author}{\bibfnamefont{G.}~\bibnamefont{Xu}},
\bibinfo{author}{\bibfnamefont{X.}~\bibnamefont{Dai}}
\bibnamefont{and} \bibinfo{author}{\bibfnamefont{Z.}~\bibnamefont{Fang}}
\bibinfo{journal}{Chin. Phys. Lett.}
\textbf{\bibinfo{volume}{26}}, \bibinfo{pages}{017401}
(\bibinfo{year}{2009}).

\bibitem[{\citenamefont{Borisenko et~al.}(2010)\citenamefont{Borisenko, Zabolotnyy, Evtushinsky, Kim, Morozov, Yaresko, Kordyuk, Behr, Vasiliev, Follath, B\(\ddot{\text{u}}\)chner}}]{Borisenko2010}
\bibinfo{author}{\bibfnamefont{S.}~\bibnamefont{Borisenko}},
\bibinfo{author}{\bibfnamefont{V.}~\bibnamefont{Zabolotnyy}},
\bibinfo{author}{\bibfnamefont{D.}~\bibnamefont{Evtushinsky}},
\bibinfo{author}{\bibfnamefont{T.}~\bibnamefont{Kim}},
\bibinfo{author}{\bibfnamefont{I.}~\bibnamefont{Morozov}},
\bibinfo{author}{\bibfnamefont{A.}~\bibnamefont{Yaresko}},
\bibinfo{author}{\bibfnamefont{A.}~\bibnamefont{Kordyuk}},
\bibinfo{author}{\bibfnamefont{G.}~\bibnamefont{Behr}},
\bibinfo{author}{\bibfnamefont{A.}~\bibnamefont{Vasiliev}},
\bibinfo{author}{\bibfnamefont{R.}~\bibnamefont{Follath}}
\bibnamefont{and} \bibinfo{author}{\bibfnamefont{B.}~\bibnamefont{B\(\ddot{\text{u}}\)chner}},
\bibinfo{journal}{Phys. Rev. Lett.}
\textbf{\bibinfo{volume}{105}}, \bibinfo{pages}{067002}
(\bibinfo{year}{2010}).


\bibitem[{\citenamefont{Eschrig et~al.}(2009)\citenamefont{Eschrig, Koepernik}}]{Eschrig2009}
\bibinfo{author}{\bibfnamefont{H.}~\bibnamefont{Eschrig}}
\bibnamefont{and} \bibinfo{author}{\bibfnamefont{K.}~\bibnamefont{Koepernik}}
\bibinfo{journal}{Phys. Rev. B}
\textbf{\bibinfo{volume}{80}}, \bibinfo{pages}{104503}
(\bibinfo{year}{2009}).

\bibitem[{\citenamefont{Lankau et~al.}(2010)\citenamefont{Lankau, Koepernik, Borisenko, Zabolotnyy, B\(\ddot{\text{u}}\)chner, van den Brink, Eschrig}}]{Lankau2010}
\bibinfo{author}{\bibfnamefont{A.}~\bibnamefont{Lankau}},
\bibinfo{author}{\bibfnamefont{K.}~\bibnamefont{Koepernik}},
\bibinfo{author}{\bibfnamefont{S.}~\bibnamefont{Borisenko}},
\bibinfo{author}{\bibfnamefont{V.}~\bibnamefont{Zabolotnyy}},
\bibinfo{author}{\bibfnamefont{J.}~\bibnamefont{van den Brink}}
\bibnamefont{and} \bibinfo{author}{\bibfnamefont{H.}~\bibnamefont{Eschrig}}
\bibinfo{journal}{Phys. Rev. B}
\textbf{\bibinfo{volume}{82}}, \bibinfo{pages}{184518}
(\bibinfo{year}{2010}).

\bibitem[{\citenamefont{Nozieres et~al.}(1999)\citenamefont{Nozieres, Pistolesi}}]{Nozieres1999}
\bibinfo{author}{\bibfnamefont{P.}~\bibnamefont{Nozi\(\grave{\text{e}}\)res}}
\bibnamefont{and} \bibinfo{author}{\bibfnamefont{F.}~\bibnamefont{Pistolesi}},
\bibinfo{journal}{Eur. Phys. J. B}
\textbf{\bibinfo{volume}{10}}, \bibinfo{pages}{649-662}
(\bibinfo{year}{1999}).

\bibitem[{\citenamefont{Brydon et~al.}(2011)\citenamefont{Brydon, Daghofer, Timm, Brink}}]{Brydon2011}
\bibinfo{author}{\bibfnamefont{P.~M.~R.}~\bibnamefont{Brydon}},
\bibinfo{author}{\bibfnamefont{M.}~\bibnamefont{Daghofer}},
\bibinfo{author}{\bibfnamefont{C.}~\bibnamefont{Timm}},
\bibnamefont{and} \bibinfo{author}{\bibfnamefont{J.}~\bibnamefont{van den Brink}},
\bibinfo{journal}{Phys. Rev. B}
\textbf{\bibinfo{volume}{83}}, \bibinfo{pages}{060501}
(\bibinfo{year}{2011}).

\bibitem{Co1} Tom Berlijn, Chia-Hui Lin, William Garber, and Wei Ku, Phys. Rev. Lett. \textbf{108}, 207003 (2012)
\bibitem{Co2} S. Ideta, T. Yoshida, I. Nishi, A. Fujimori, Y. Kotani, K. Ono, Y. Nakashima, S. Yamaichi, T. Sasagawa,
  M. Nakajima, K. Kihou, Y. Tomioka, C. H. Lee, A. Iyo, H. Eisaki, T. Ito, S. Uchida, and R. Arita,
  Phys. Rev. Lett. \textbf{110}, 107007 (2013)

\bibitem{JPHu} Jiangping Hu and Hong Ding, Scientific Reports \textbf{2}, 381 (2012)

\end{thebibliography}
\end{document}